\documentclass[aps,prl,floatfix,twocolumn,tightenlines,amsmath,amssymb]{revtex4}

\usepackage{graphicx}
\usepackage{amssymb}
\usepackage{color}
\usepackage{epstopdf}
\usepackage{tipa}
\usepackage{hyperref}

\newcommand{\bra}[1] {\langle #1 |}
\newcommand{\ket}[1] {| #1 \rangle}
\newcommand{\braket}[2] {\langle #1 | #2 \rangle}
\newcommand{\ketbra}[2] {| #1\rangle \langle  #2 |}

 {\begin{list}{}%
         {\setlength{\leftmargin}{#1} \setlength{\rightmargin}{#2}}%
         \item[]%
 }
 {\end{list}}

\begin{document}

\title{Observation of Entanglement-Dependent Two-Particle Holonomic Phase}
\author{J. C. Loredo\footnote{Electronic address: {juan.loredo1@gmail.com}}, M. A. Broome, D. H. Smith and A. G. White}
\affiliation{Centre for Engineered Quantum Systems, Centre for Quantum Computer and Communication Technology, and School of Mathematics and Physics, University of Queensland, Brisbane, QLD 4072, Australia.}

\begin{abstract}
{Holonomic phases---geometric and topological---have long been an intriguing aspect of physics. They are ubiquitous, ranging from observations in particle physics to applications in fault tolerant quantum computing. However, their exploration in particles sharing genuine quantum correlations lack in observations. Here we experimentally demonstrate the holonomic phase of two entangled-photons evolving locally, which nevertheless gives rise to an entanglement-dependent phase. We observe its transition from geometric to topological as the entanglement between the particles is tuned from zero to maximal, and find this phase to behave more resilient to evolution changes with increasing entanglement. Furthermore, we theoretically show that holonomic phases can directly quantify the amount of quantum correlations between the two particles. Our results open up a new avenue for observations of holonomic phenomena in multi-particle entangled quantum systems.}
\end{abstract}

\maketitle
In differential geometry, holonomy accounts for the difference between a parallel-transported vector  along a geodesic---i.e. shortest path---and any other curve. It is a direct manifestation of the geometry \emph{and} topology of a given curved space. A physical system evolving in its own multi-dimensional parameter space will exhibit holonomies as a result of these geometric and topological structures. Consequently, holonomies have physical manifestations, ranging from Thomas precession to the Aharonov-Bohm effect.

In quantum systems, the holonomy manifests as a phase imparted on the wavefunction~\cite{HolAndBerry:Simon}. When the quantum parameter space is \emph{simply} connected, holonomies are continuous-valued with respect to continuous deformations of the trajectory. These are geometric phases~\cite{Berry:Quantal} and they depend on the space's curvature. Conversely, when the parameter space is not simply-connected discrete-valued topological phases appear~\cite{Milman:TP2q,Liming:SO3MES}. We refer to both geometric and topological as holonomic phases.

Holonomies are of fundamental interest and have important applications, for example, in holonomic quantum computation~\cite{Zanardi:HQC,Cirac:GQC,Lidar:HQC,Sjoqvist:NAHQC}, where matrix-valued geometric phase transformations play the role of quantum logic gates. This scheme has received a great deal of attention due to its potential to overcome decoherence~\cite{Filipp:ExpRobustBerry}, and has recently been experimentally realized in different architectures~\cite{ExpNaNaHQC:Filipp2,ExpNAHQC:Feng}. 

In the quantum regime, holonomic phases have been observed in particles encoding one qubit~\cite{Kwiat:GPSP,Ericsson:ExpMGP,Cucchietti:ExpGPnU}, as well as two particle systems encoding uncorrelated two-qubit states~\cite{kobayashi:NLGP}. In addition, topological phases have been observed in classical systems emulating the behaviour of entanglement, for example, so-called non-separable states between the polarization and transverse modes of a laser~\cite{Khoury:ExpTP}, or pseudo-entanglement in NMR~\cite{Du:ExpTPNRM}. Lacking up to now, however, is the exploration of holonomic phases between genuinely entangled quantum particles. 

Here we demonstrate both geometric and topological phases appearing in the joint wavefunction of two separate, and genuinely-entangled, particles whose correlations can be tuned from vanishing to maximal. 
 
\begin{figure*}[htp]
	\includegraphics[width=0.85\textwidth]{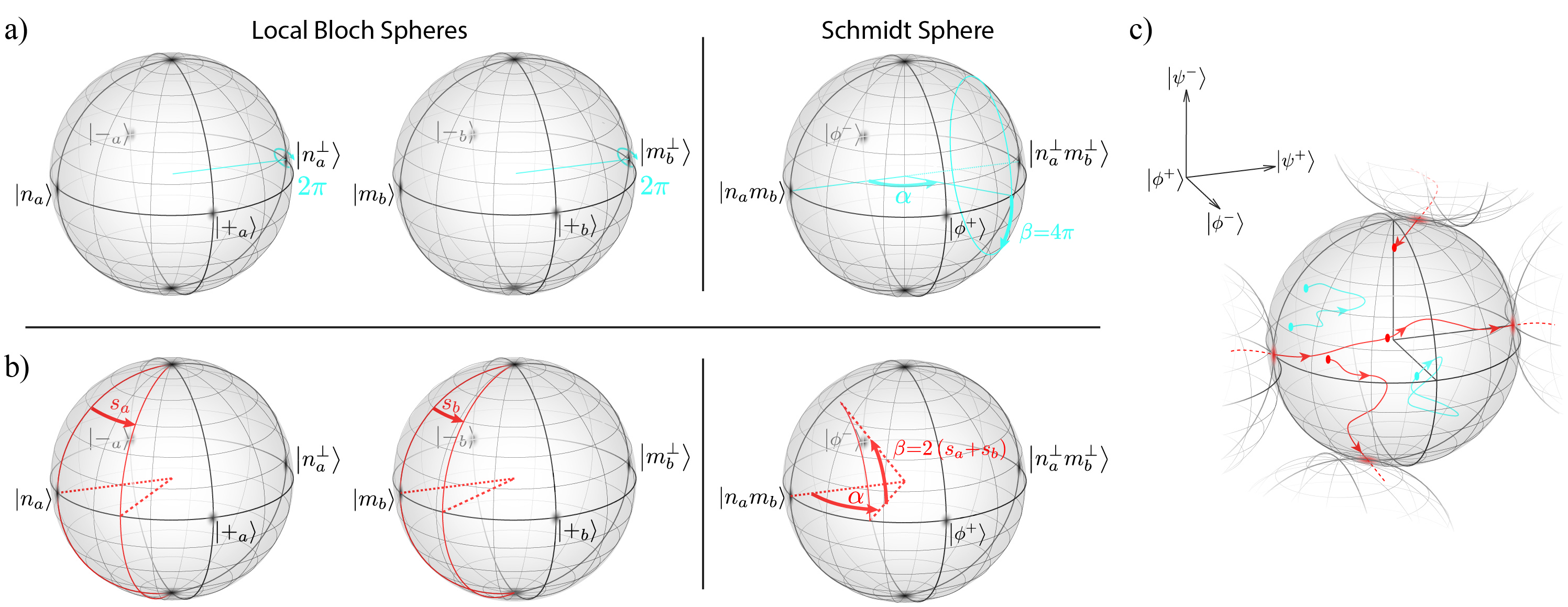}
	\caption{\label{fig:fig1}a) A cyclic Schmidt evolution with $\theta_{\textrm{S}}{=}2\pi$. Left and middle spheres show trajectories traced out in the local Bloch spheres of each qubit defined in the Schmidt basis. Here, they evolve at a single point and therefore enclose no area, leading to no gain in holonomic phase. The right sphere represents the trajectory spanned by the evolution of $\alpha$ and $\beta$ in the Schmidt sphere. In contrast to the local trajectories, a holonomic phase still arises as a result of the area enclosed by this trajectory. Equivalent Schmidt evolutions of non-entangled states induce a zero holonomic phase. b) In contrast, segmented evolutions defined by angles $s_a$ and $s_b$ induce holonomies appearing from both local and Schmidt spheres. The parameters $s_a$ and $s_b$ define the opening angles for the projected curves of $\rho_a$ and $\rho_b$ onto the local Bloch spheres. In the right sphere, a rotation angle $2(s_a{+}s_b)=\beta$ around $\ket{n_a m_b}$ also contributes to the holonomy. Equivalent evolutions of non-entangled states induce a non-zero holonomic phase. c) Depiction of the double-connected parameter space of maximally entangled states: $\mathrm{SO}(3)$ in $\mathbb{R}^3$ with a border at $S^2_{\pi}$. This border is a $2$-sphere of radius $\pi$ with identified antipodal points. Blue trajectories represent arbitrary  evolutions for one homotopy-class along which no phase is gained. Red curves represent evolutions of the other homotopy-class after which a $\pi$ phase appears on the wave-function.}
	\end{figure*}

To better elucidate two-qubit holonomic phases, consider an arbitrary two-qubit pure state written in its Schmidt decomposition:
	\begin{equation}
		\ket{\psi(0)}=e^{-i\beta/2}\cos\frac{\alpha}{2}\ket{n_a m_b}+e^{i\beta/2}\sin\frac{\alpha}{2}\ket{n_a^\perp m_b^\perp},
		\label{eq:purestate}
	\end{equation}
where $\alpha\in[0,\pi]$ and $\beta\in[0,2\pi]$ parametrise the Schmidt sphere~\cite{Sjoqvist:GPentPair} of a correlation space, and $\ket{n_a m_b} $, $\ket{n_a^\perp m_b^\perp}$ are orthogonal product states defining the Schmidt basis, see Fig.~\ref{fig:fig1}a, b. While a treatment of the two-qubit space can be carried out formally~\cite{jakob:blochvectors}, it is more intuitive to represent evolutions in it with the trajectories that the reduced density matrices, $\rho_a$ and $\rho_b$, undertake on their corresponding local Bloch spheres together with the curve that $\alpha$ and $\beta$ project onto the Schmidt sphere. 

For a non-maximally entangled state, $\ket{\psi(0)}$ has preferred directions, given by unit vectors $\hat{a}$ and $\hat{b}$, on each qubit's Bloch sphere. That is, the reduced density matrix of, say, system $a$ is given by $\rho_a{=}\textrm{Tr}_b(\ketbra{\psi}{\psi}){=}\tfrac{1}{2}\left(\mathbb{I}+\cos{\alpha} {\,} \hat{a}\cdot\vec{\sigma}\right)$, where $\vec{\sigma}{=}\left(\sigma_x,\sigma_y,\sigma_z\right)$ denote the Pauli matrices. Accordingly, $\ket{\psi(0)}$ spans a six-dimensional parameter space.

From the state $\ket{\psi(0)}$, an entanglement-induced holonomic phase will appear as a result of the special ``Schmidt evolution". We define a Schmidt evolution as a bi-local rotation of $\theta_{\textrm{S}}$, of both qubits around their preferential directions $\hat{a}$ and $\hat{b}$, see Fig.~\ref{fig:fig1}a. The holonomic phase of this evolution is calculated---in the standard way---as a difference~\cite{Mukunda:KinAppGP} 
	\begin{equation}
		\Phi_\textrm h=\Phi_\textrm{P}-\Phi_\textrm{dyn},
		\label{eq:holonomicphase}
	\end{equation}
with $\Phi_\textrm{P}{=}\arg\braket{\psi(0)}{\psi(\tau)}$ the Pancharatnam~\cite{Panch:Phase} and $\Phi_\textrm{dyn}{=}\textrm{Im}\int_0^\tau\braket{\psi(t)}{ {\dot\psi}(t)}dt$ the dynamical phase, and $\ket{\psi(t)}, t\in[0,\tau]$, denotes the evolving state.

While usually $\Phi_\textrm{h}$ is regarded as a geometric phase only~\cite{Mukunda:KinAppGP}, its value arises from both the geometry (curvature) \emph{and} topology (connectedness) of the parameter space. It has therefore become more routine to identify parts of $\Phi_\textrm{h}$ as being of either  geometric or topological origin~\cite{Milman:PhDyn}. For instance, maximally-entangled two-qubit pure states (MES) can only induce a phase of topological origin regardless of their evolution~\cite{Milman:TP2q,Liming:SO3MES,Milman:PhDyn}.

Canonically, the amount of entanglement in a two-qubit state can be measured by the tangle $\mathbb{T}$ (concurrence squared \cite{Wooters:Conc}). In a pure system as given in Eq.~\ref{eq:purestate}, it is determined by the relative populations of the Schmidt basis: $\mathbb{T}=\sin^2\alpha$, and ranges from $0$ for separable states up to $1$ for maximally-entangled states. Consequently, a Schmidt evolution will give rise to, see Supplemental Material, an entanglement-induced holonomic phase given by:
	\begin{equation}\label{eqhpent}
			\Phi_\textrm{h}^{\textrm{ent}}=\arg\left( \cos\theta_{\textrm{S}}{-}i\sqrt{1{-}\mathbb{T}}\sin\theta_{\textrm{S}} \right){+}\theta_{\textrm{S}}\sqrt{1{-}\mathbb{T}}.
	\end{equation}
Importantly, $\Phi_{\textrm h}^{\textrm{ent}}$ behaves monotonically with the amount of entanglement, measuring 0 for separable states and its maximum for MES (value depended on $\theta_{\textrm{S}}$). For instance, for the evolution depicted in Fig.~\ref{fig:fig1}a, $\theta_\mathrm{S}{=}2\pi$, and $\Phi_\textrm{h}^\textrm{ent}=-2\pi\left(1-\sqrt{1-\mathbb{T}}\right)$. While there are extensive theoretical studies of holonomic phases in mixed, as well as pure, entangled systems~\cite{Tong:GPentSub,Ericsson:MGPentLocal,Johansson:TPmultiEnt}, it remains an open question as to whether a holonomic phase quantifying entanglement can be found for mixed states.

	\begin{figure*}[htp!]
	\includegraphics[width=0.9\textwidth]{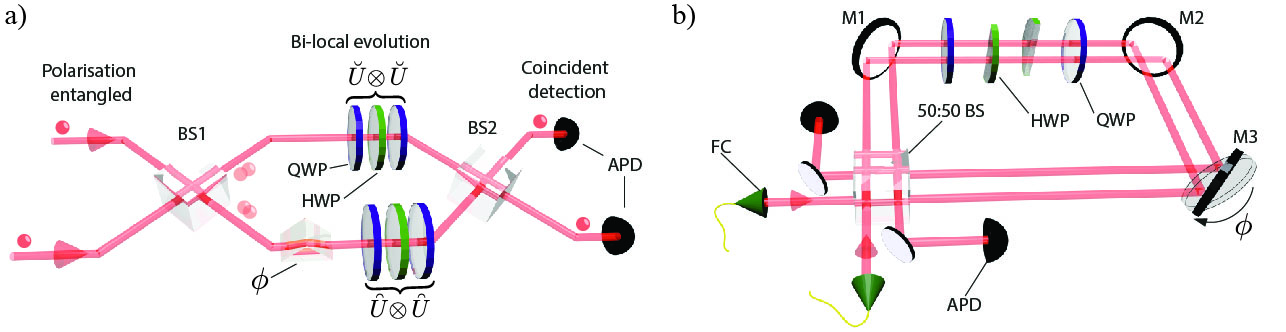}
	\caption{\label{fig:fig3}a)~Depiction of the experimental method. Two indistinguishable photons, encoding a two-qubit entangled state $\ket{\psi}$ in their polarization, pass through an interferometer and are detected in coincidence with avalanche-photo-diodes (APD) at the output. Due to the Hong-Ou-Mandel effect at the first beam-splitter BS1, a coincidence signal $c$ after the exit of the interferometer through BS2, cannot distinguish between different paths travelled in the interferometer. The resulting  holonomic phase gained by the evolved state $\ket{\psi'}{=}\textit{\textroundcap{U}}^\dagger\breve{U}\otimes\textit{\textroundcap{U}}^\dagger\breve{U}\ket{\psi}$, as well as the phase $\phi$, thus appear as a modulation in the coincidence given by $c\propto|e^{2i\phi}\ket{\psi}+\ket{\psi'} |^2$. b)~Experimental implementation of the method. Single photons are injected via single-mode fibre couplers (FC) into a displaced-Sagnac interferometer configuration, composed of one $50{:}50$ beam-splitter (BS) and three mirrors M1-3. Polarisation evolution is performed using two common-path quarter wave-plates QWP and two semi-circular half wave-plates HWP in separate optical paths. Mirror M3 is rotated via a micro-translation stage to control $\phi$.} 
	\end{figure*}

Recalling that experimentally it is the total Pancharatnam phase that is observed~\cite{Loredo:Panch}, we can choose evolutions for which the dynamical component vanishes, ensuring the total phase gained is holonomic in nature only. We achieve this using the the bi-local segmented evolutions characterised by the opening angles $s_a$ and $s_b$ on qubits $a$ and $b$ respectively, see Fig.~\ref{fig:fig1}b. These trajectories are connected geodesics, meaning any dynamical phase is identically zero. However, not being Schmidt evolutions, the holonomic phase arises from trajectories in \emph{both} local and Schmidt spheres, but importantly remains monotonic with entanglement:
	\begin{equation}\label{eqhpexp}
		\Phi_\textrm{h}=\mp\arctan\left( \sqrt{1-\mathbb{T}}\tan(2s) \right),		
	\end{equation}
where the sign is $-$ ($+$) if the joint state is more populated in $\ket{n_am_b}$ ($\ket{n_a^{\perp}m_b^{\perp}}$) of the Schmidt basis, see Supplemental Material, and $s$ defines the evolution undertaken by $\ket{\psi(0)}$. In Eq.~(\ref{eqhpexp}), the opening angles $s_a{=}s_b{=}s$.

One important feature of the state in Eq.~(\ref{eq:purestate}) is the change that occurs to the parameter space as a result of increasing tangle. As $\mathbb{T}{\rightarrow}1$, previously separated states in the two-qubit parameter space become less distinguishable, and eventually some become identical at $\mathbb{T}{=}1$. At this point the parameter space collapses from six to three dimensions represented by the double-connected $\mathrm{SO}(3)$ ball \cite{Milman:TP2q, Liming:SO3MES}, see Fig.~\ref{fig:fig1}c. Spaces of this kind---not simply connected---allow state trajectories that are topologically distinct, i.e. cannot be continuously transformed into one another.

Trajectories in the $\mathrm{SO}(3)$ ball are classified by two different homotopy-class families: those that cross the border $S^2_{\pi}$---a 2-sphere of radius $\pi$---an odd number of times and those crossing it an even number of times (zero included). Physically, crossing $S^2_{\pi}$ $l$-times results in a $l\pi$ phase on the wavefunction. For instance, if $\alpha{=}\pi/2$ in the Schmidt evolution shown in Fig.~\ref{fig:fig1}a, then its trajectory in $\mathrm{SO}(3)$ crosses $S^2_{\pi}$ twice, picking up a $2\pi$ topological phase.

In order to observe the entanglement-dependent holonomic phase given by Eq.~(\ref{eqhpexp}), we implement a method, depicted in Fig.~\ref{fig:fig3}, that works as follows: We generate a two-qubit state, $\ket{\psi}$, in the polarization of two single-photons whose tangle can be tuned from $\mathbb{T}{=}0{\rightarrow}1$~\cite{Aggie2011,Smith:CQS}. Upon meeting at the first $50{:}50$ beam-splitter BS$1$ of the interferometer depicted in Fig.~\ref{fig:fig3}a, the photons are subject to non-classical interference~\cite{mandel:HOM}, after which they exit via the same spatial mode. Regardless of the specific form of $\ket{\psi}$, photon bunching can always be achieved by engineering other degrees of freedom if necessary \cite{Monken:MultiHOM}. Consequently, the joint state of the system, $\ket{\psi}$, remains in either of the two paths of the interferometer, whose optical path lengths are equal. The joint state then undergoes a polarization evolution, composed of two auxiliary evolutions $\breve{U}$ and $\textit{\textroundcap{U}}$ in separate paths of the interferometer. 

The information about which-path the photons followed is erased by a second $50{:}50$ beam-splitter BS$2$, after which point the photons are detected. As such, the two-photon coincidence signal, $c$, will exhibit interference behaviour modulated by all relative phases between the two arms of the interferometer.  That is, caused by a physical optical path-difference phase $\phi$ and the Pancharatnam phase $\Phi_\textrm{P}$ arising from the polarization evolutions.

We prepared the initial polarization-entangled state $\ket{\psi}{=}\cos\frac{\alpha}{2}\ket{HH}+\sin\frac{\alpha}{2}\ket{VV}$, where $\alpha$ is a tuneable parameter defining the amount of entanglement $\mathbb{T}{=}\sin^2\alpha$, and $\ket{H}$ and $\ket{V}$ denote the horizontal and vertical polarizations. The state is injected into a displaced-Sagnac interferometer configuration, shown in Fig.~\ref{fig:fig3}b, where no active path locking is required. Using the notation $\sigma_1{=}\ket{H}\bra{H}-\ket{V}\bra{V}$ and $\sigma_2{=}\ket{H}\bra{V}+\ket{V}\bra{H}$, we choose an evolution
	\begin{equation}
		U\equiv\textit{\textroundcap{U}}^\dagger\breve{U}=e^{i\frac{\pi}{4}\sigma_2} e^{-i\frac{\pi}{2}\left(\sigma_1\sin{s}+\sigma_2\cos{s}\right)} e^{i\frac{\pi}{4}\sigma_2},
		\label{eq:U}
	\end{equation}
where $s$ equals the opening angles of the corresponding geodesic trajectories in the local spaces, see Fig.~\ref{fig:fig1}b. Since the evolutions $\breve{U}$ and $\textit{\textroundcap{U}}$ are chosen to induce no dynamical phase, from Eq.~\ref{eq:holonomicphase} the Pancharatnam phase matches the holonomic phase. Thus by controlling the variable phase $\phi$---introduced by slightly rotating the mirror M3 of the interferometer, see Fig.~\ref{fig:fig3}b---the two-photon coincidence will modulate as:
	\begin{eqnarray}\label{eqc}
		c&=& \frac{1}{4}\left|e^{2i\phi}\ket{\psi}+(\textit{\textroundcap{U}}^\dagger\breve{U}{\otimes}\textit{\textroundcap{U}}^\dagger\breve{U})\ket{\psi} \right|^2 \nonumber \\
		&=&\frac{1}{2}\left[ 1+v\cos(2\phi-\Phi_\textrm{h}) \right],
	\end{eqnarray}
where $v{=}|\braket{\psi}{\psi'}|$ is the interference visibility and $\Phi_\textrm{h}{=}\arg\braket{\psi}{\psi'}$ is the holnomic phase gained during the effective evolution $\ket{\psi'}{=}U{\otimes}U \ket{\psi}$. Consequently, we can determine $\Phi_{\textrm{h}}$ by measuring coincidence modulation as a function of $\phi$, see Supplemental Material.

	\begin{figure*}[htb!]
	\includegraphics[width=170mm]{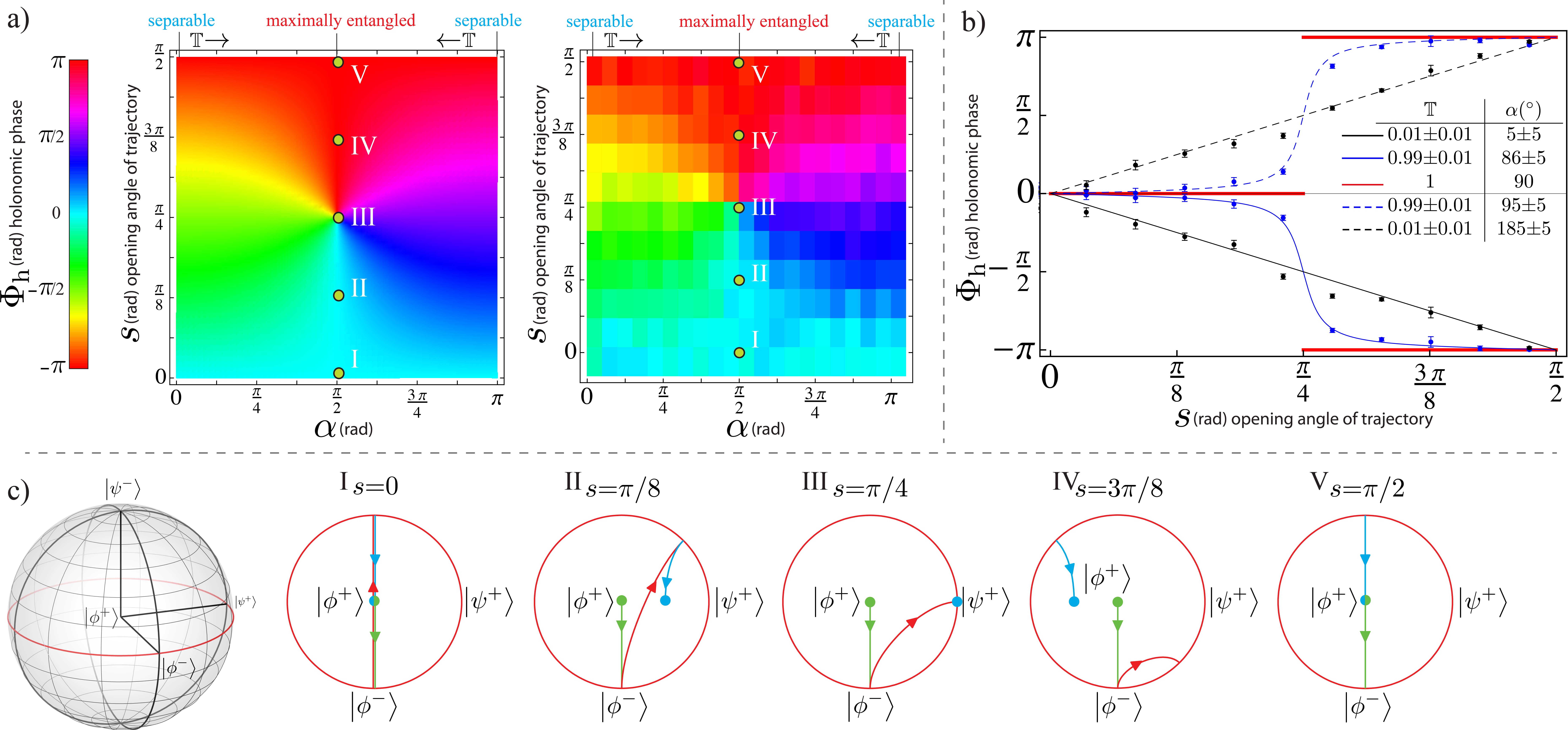}
	\caption{\label{fig:fig4}Experimental results. a) In a cyclic colour-scale, predicted (left) and measured (right) results for ${\Phi_\textrm{h}{=}-\arctan\left(\cos(\alpha)\tan(2s) \right)}$ (equivalent to Eq.~(\ref{eqhpexp})). Centres of the rectangular measured data blocks represent the corresponding $(\alpha,s)$ coordinates. b) The black data points and theoretical curves correspond to measurements of holonomic phases with an initial state $\ket{HH}$ (solid curve) and $\ket{VV}$ (dashed curve) for a fitted level of tangle, $\mathbb{T}{=}0.01{\pm}0.01$. Blue data points and theoretical curves show the extent to which we observe the topological behaviour of $\Phi_{\textrm{h}}$ for a highly entangled state. Here the data is fitted to a state tangle of $\mathbb{T}{=}0.99{\pm}0.01$, for the cases in which $\ket{\psi(0)}$ is more populated by $\ket{HH}$ ($\alpha{=}\pi/2{-}\epsilon$) shown by the solid curve, and $\ket{VV}$ ($\alpha{=}\pi/2{+}\epsilon$) given by the dashed curve, for $0{<}\epsilon{\ll}1$. The red curve corresponds to the theoretical ideal case of $\mathbb{T}{=}1$. Errors are calculated via Poissonian counting statistics. c) Depiction of cyclic (I,V) and non-cyclic (II,III,IV) evolutions of MES in a plane of the double-connected $\mathrm{SO}(3)$ ball. Green, red and blue coloured-curves denote the first, second and third part of each evolution, respectively. Evolution III for which $s{=}\pi/4$, marks the switch between two distinct homotopy-classes, those that cross the $S^2_\pi$ border zero times (I and II) and one time (IV and V).
	} 
	\end{figure*}

The polarization evolution in Eq.~\ref{eq:U} is implemented using quarter- $Q$ and half-wave-plates $H$ in the arrangement $U{=}Q(-\frac{\pi}{4})H(\frac{\pi}{4}{-}\frac{s}{2})Q(-\frac{\pi}{4})$, where the arguments inside parentheses indicate the angle of the wave-plate optic axis in the laboratory frame. This unitary evolution is built from auxiliaries in separate arms of the interferometer, given by $\breve U{=}Q(-\frac{\pi}{4})H(\frac{\pi}{4}{-}\frac{s}{4})Q(-\frac{\pi}{4})$ and $\textit{\textroundcap{U}}{=}Q(\frac{\pi}{4})H(-\frac{\pi}{4}{-}\frac{s}{4})Q(\frac{\pi}{4})$. Physically, we perform these evolutions using two common-path quarter-wave-plates with angles fixed at ${-}\frac{\pi}{4}$ and two semi-circular half-wave-plates, one in each path, see Fig.~\ref{fig:fig3}b.
 
Figure~\ref{fig:fig4}a shows our predicted and measured results for $\Phi_\textrm{h}$, as a function of entanglement represented by $\alpha$, and the opening angle $s$ of the trajectories. The solid and dashed black lines in Fig.~\ref{fig:fig4}b show theoretical predictions for the bipartite holonomic phase for a fitted value of tangle, $\mathbb{T}{=}0.01\pm0.01$. In this case the holonomy is simply the sum of individual geometric phases $\Phi_\textrm{h}^\text{sep}{=}\mp(s{+}s)$. 

Conversely, as the amount of entanglement increases to a maximum, i.e. as $\alpha{\to}\pi{/}2$, the holonomic phase becomes less affected by changes in the trajectory angle $s$. Instead, two attractors at $\Phi_\textrm{h}{=}0$ and $\Phi_\textrm{h}{=}\pi$ appear, becoming its only possible values when $\mathbb{T}{=}1$, shown by the solid-red curves in Fig.~\ref{fig:fig4}b. Thus, the tuning from $\alpha{=}0\to\alpha{=}\pi/2$ results in a geometric-to-topological transition of the holonomic phase. The solid and dashed blue curves in Fig.~\ref{fig:fig4}b show our measurements for the fitted value of tangle, $\mathbb{T}{=}0.99\pm0.01$.

This reported tangle is the best-fitted value to our data using Eq.~(\ref{eqhpexp}). To confirm the presence of a high amount of entanglement, we carried out a full quantum state tomography of the two-photon state, resulting in a tangle of $\mathbb{T}_\text{tomo}{=}0.959\pm0.001$, see Supplemental Material.

As discussed previously, when $\mathbb{T}{=}1$, the corresponding evolutions follow trajectories in the double-connected $\mathrm{SO}(3)$ space, thus giving rise to two topologically distinct paths. Figure~\ref{fig:fig4}c shows the corresponding paths for five such evolutions, for which the trajectory with $s{=}\frac{\pi}{4}$ represents the switch between two the distinct homotopy-classes. 
	
Experiments~\cite{Khoury:ExpTP, Du:ExpTPNRM} observing such topological phases have been realised with cyclic evolutions using classical states that are formally equivalent to a MES, corresponding to the top and bottom data points at $\alpha{=}\pi/2$ in Fig.~\ref{fig:fig4}a. In contrast, we additionally demonstrated topological behaviour for explicitly non-cyclic evolutions, see Fig.~\ref{fig:fig4}c, of genuinely entangled quantum systems.

We experimentally demonstrated that the wavefunction of a pair of qubits picks up a phase factor of holonomic nature in both cyclic and non-cyclic evolutions. In contrast to conventional measurements of single-qubit holonomic phases~\cite{Kwiat:GPSP, Ericsson:ExpMGP}, the phase shift observed in our work is dependent on genuine quantum correlations. We find that the holonomic phase becomes more resilient to evolution changes with increasing entanglement, which indicates that quantum correlations can be utilised to enhance holonomic robustness and there may be advantages in using not only geometric but the full range of holonomic phases. Naturally, this leads to the question as to whether more general forms of quantum correlations---most notably discord~\cite{Ollivier:QDiscord}---could be the underlying reason for this enhancement. 

Finally, we derived an entanglement-induced holonomic phase that can be used to quantify the amount of quantum correlations between a pair of pure-state qubits. This result provides a measurable quantity arising solely from entanglement and it is a step in gaining a broader understanding of the geometric interpretation of quantum correlations. We expect that this work will strongly motivate new proposals for more robust holonomic quantum computation and trigger observations of holonomies in multi-partite entangled states of qubits or higher-dimensional qudit systems.

\subsection{Acknowledgments}

We thank G. J. Milburn for stimulating discussions, and A. Fedrizzi, C. Branciard, I. Kassal and M. Ringbauer for critical feedback.  This work was supported by: The ARC Centre for Quantum Computation and Communication Technology (CE110001027), Centre for Engineered Quantum Systems (CE110001013), and Federation Fellow program (FF0668810).

%%%%%%%%%%%%%%%%%%%%%%%%%%%%%%%%%%%%%%%%

\subsection{Supplemental Material}

\subsection{I. Calculation of holonomic phase of entanglement}

Consider the arbitrary initial two-qubit pure state written in its Schmidt decomposition,
	\begin{equation}
		\ket{\psi(0)}=\cos\frac{\alpha}{2}e^{-i\beta/2}\ket{n_a m_b}+\sin\frac{\alpha}{2}e^{i\beta/2}\ket{n_a^\perp m_b^\perp},
	\end{equation}
which is then subject to a rotation of both qubits around their preferential directions $\hat{a}$ and $ \hat{b}$ in their corresponding local Bloch spheres. That is, a bi-local rotation $U^{\mathcal S}(\theta){=}U_{a}^{\mathcal S(n)}(\theta)\otimes U_b^{\mathcal S(m)}(\theta)$ such that $U_j^{\mathcal S(k)}(\theta)=\exp\left(-i\theta\sigma_j^{(k)}/2\right)$, where $\theta$ is the angle of rotation, $\sigma_j^{(k)}=\eta\ket{k_j}\bra{k_j}-\eta\ket{k_j^\perp}\bra{k_j^\perp}$,  and $\eta$ is the sign of $\cos\alpha$. The holonomic phase of such evolution can be calculated from the formula
	\begin{equation}
		\Phi_\textrm h=\arg\braket{\psi(0)}{\psi(\theta)}-\textrm{Im}\int_0^\theta\braket{\psi(\theta')}{ {\dot\psi}(\theta')}d\theta',
	\end{equation}
where $\Phi_\textrm{P}{=}\arg\braket{\psi(0)}{\psi(\theta)}$ and $\Phi_\textrm{dyn}{=}\textrm{Im}\int_0^\theta\braket{\psi(\theta')}{ {\dot\psi}(\theta')}d\theta'$ are the Pancharatnam and dynamical phases respectively. This gives,
	\begin{eqnarray}\nonumber
		\arg\braket{\psi(0)}{\psi(\theta)}&=&\arg\left( \cos\theta-i\eta\cos\alpha\sin\theta \right)\\ 
		&=&\arg\left( \cos\theta-i|\cos\alpha|\sin\theta \right),
	\end{eqnarray}
and
	\begin{eqnarray}\nonumber
		\textrm{Im}\int_0^\theta\braket{\psi(\theta')}{ {\dot\psi}(\theta')}d\theta'&=&\textrm{Im}\int_0^\theta -i\eta\cos\alpha d\theta'\\ 
		&=&-|\cos\alpha|\theta.
	\end{eqnarray}
Using the expression for the tangle $\mathbb{T}{=}\sin^2\alpha$, we find the holonomic phase of entanglement to be
	\begin{eqnarray}
		\Phi_\textrm{h}^{\textrm{ent}}=\arg\left( \cos\theta{-}i\sqrt{1{-}\mathbb{T}}\sin\theta \right){+}\theta\sqrt{1{-}\mathbb{T}},
	\end{eqnarray}
which for  $\theta{=}2\pi$ takes the form
	\begin{equation}
		\Phi_\textrm{h}^\textrm{ent}=-2\pi\left(1-\sqrt{1-\mathbb{T}}\right).
	\end{equation}
	
\subsection{II. Holonomic phase in evolutions with vanishing dynamical phase}

In our experiment we prepare the initial polarization-entangled state $\ket{\psi}{=}\cos\frac{\alpha}{2}\ket{HH}+\sin\frac{\alpha}{2}\ket{VV}$, where $\ket{H}$ and $\ket{V}$ denote horizontal and vertical polarizations respectively. We choose bi-local evolutions $U\otimes U$, with
	\begin{equation}
		U=e^{i\frac{\pi}{4}\sigma_2} e^{-i\frac{\pi}{2}\left(\sigma_1\sin{s}+\sigma_2\cos{s}\right)} e^{i\frac{\pi}{4}\sigma_2},
	\end{equation}
where $\sigma_1{=}\ket{H}\bra{H}-\ket{V}\bra{V}$, $\sigma_2{=}\ket{H}\bra{V}+\ket{V}\bra{H}$ and $s$ equals the opening angles of the trajectories projected onto the Bloch spheres. For this particular evolution, one can demonstrate that the dynamical phase vanishes at every step of the evolution. Then the holonomic phase is calculated as
	\begin{eqnarray}\nonumber
		\Phi_\textrm{h}&=&\arg\left( \cos2s-i\cos\alpha\sin2s\right)\\ 
		&=&-\arctan\left( \cos\alpha\tan2s \right).
	\end{eqnarray}
Using $\cos\alpha{=}\eta\sqrt{1-\mathbb{T}}{=}\pm\sqrt{1-\mathbb{T}}$, where the sign is $+$ ($-$) if $\ket{HH}$  ($\ket{VV}$) is more populated, we find that
	\begin{equation}
		\Phi_\textrm{h}=\mp\arctan\left( \sqrt{1-\mathbb{T}}\tan(2s) \right).
	\end{equation}
For equally-populated states (maximally-entangled states) the sign is irrelevant and the holonomic phase $\Phi_\textrm{h}{=}\arg(\cos2s)$ is topological, being $\Phi_\textrm{h}{=}0$ for $\cos2s\geq0$, and $\Phi_\textrm{h}{=}\pi$ for $\cos2s<0$.

\subsection{III. Measurement of the coincidence signal}

	\begin{figure}[htp!]
	\includegraphics[width=80mm]{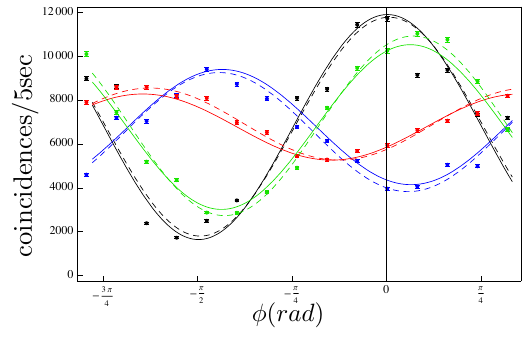}
	\caption{{Counted coincidence events.} The reference signal (solid black curve) was obtained from finding $N^i$ and $N_{0}^i$ by imposing $s^i{=}0$ and $\alpha^i{=}i\pi/20, i=0,1,...,20$ and averaging them to obtain $N{=}11911$ and $N_0{=}1616$, the phase origin, i.e. $\phi{=}0$, is set to be the average of their phases.  Black points and dashed black curve represent  the collected coincidence counts and best sinusoidal fit for $(s,\alpha){=}(0,\frac{8}{20}\pi)$. Blue, red and green points and curves correspond to the predicted experimental coincidence signal (solid curves) $c_\textrm{e}$, see Eq.~(\ref{eqce}), and best sinusoidal fits (dashed curves) for $(s,\alpha){=}(\frac{7}{10}\frac{\pi}{2},\frac{8}{20}\pi)$, $(s,\alpha){=}(\frac{6}{10}\frac{\pi}{2},\frac{11}{20}\pi)$ and $(s,\alpha){=}(\frac{3}{10}\frac{\pi}{2},\frac{12}{20}\pi)$, respectively.
	}
	\label{fig:fig1SI}
	\end{figure}

Ideally, two parameters, $s$ and $\alpha$, define the probability $c{=}\frac{1}{2}\left[ 1+v_\textrm{t}\cos(2\phi-\Phi_\textrm{h}) \right]$, with $v_\textrm{t}{=}\sqrt{(\cos2s)^2+(\cos\alpha\sin2s)^2}$ and $\Phi_\textrm{h}{=}-\arctan\left( \cos\alpha\tan2s \right)$, of detecting a coincidence event in the experimental set-up described in the main text. Then, if $N$ pairs of single-photons enter the set-up, theoretically we would observe a coincidence signal
	\begin{equation}
		c_\textrm{t}=\frac{N}{2}\left[ 1+v_\textrm{t}\cos(2\phi-\Phi_\textrm{h}) \right],
	\end{equation}
which predicts a reference, i.e. $s{=}0$, theoretical visibility $v_\textrm{t}^\textrm{r}{\equiv}\left(c^\textrm{r,max}_\textrm{t}-c^\textrm{r,min}_\textrm{t}\right){/}\left(c^\textrm{r,max}_\textrm{t}+c^\textrm{r,min}_\textrm{t}\right){=}1$, $\forall\alpha$, where $\textrm{max}$ ($\textrm{min}$) stands for maximum (minimum) value with respect to $\phi$.

	\begin{figure*}[htp!]
	\includegraphics[width=15cm]{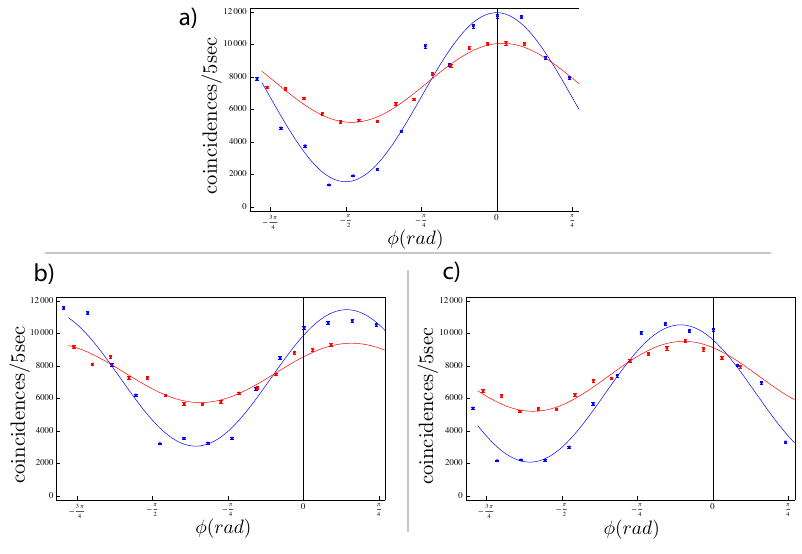}
	\caption{{Indistinguishable-distinguishable cases comparison.} Counted coincidences inside (blue dots) and outside (red dots) the Hong-Ou-Mandel dip for a) $(s,\alpha){=}(0,\frac{9}{20}\pi)$, b) $(s,\alpha){=}(\frac{4}{10}\frac{\pi}{2},\frac{14}{20}\pi)$ and c) $(s,\alpha){=}(\frac{3}{10}\frac{\pi}{2},\frac{3}{20}\pi)$. Curves represent best sinusoidal fits for the corresponding data. Inside and outside the dip we obtain $N{=}11911$, $N_0{=}1616$, and $N^\textrm{out}{=}10000$, $N^\textrm{out}_0{=}5068$, respectively. Which, from Eq.~(\ref{eqvis}), predicts an outside/inside visibility ratio of $0.43$. The observed visibility ratios are $0.41$, $0.42$ and $0.44$ for a), b) and c), respectively.
	}
	\label{fig:fig2SI}
	\end{figure*}
Experimentally, however, we observe the reference counts dropping to a minimum of non-zero $N_0$ events, which decreases the signal visibility. Imperfect $50{:}50$ beamsplitter reflection{:}transmission ratio, and higher-order photon pairs---mainly present in pulsed-pumped down-conversion sources---can decrease visibilities in interference experiments. However, in our case the former is measured to be $46{:}54$ at the source-wavelength ($820$nm) and the latter are negligible as ours is a cw-pumped down-conversion source. Accordingly, these issues have a minor contribution on the reduced visibility. On the other hand, if fully distinguishable photons (outside the Hong-Ou-Mandel dip) enter our set-up, one can show that the reference interference probability oscillates between $1$ and $1{/}2$, giving a visibility of $1{/}3$, which is in agreement with our observations outside the dip, see Fig.~{\ref{fig:fig2SI}}a. Thus, the degree of photon-distinguishability plays an important role and accounts for the reduced visibility, which in our experiment we believe arises from a temporal mode mismatch between the single photons.

The maximum and minimum number of experimental reference counts, $c_\textrm{e}^\textrm{r,max}{=}N, c_\textrm{e}^\textrm{r,min}{=}N_0$, give an experimental reference visibility $v_\textrm{e}^\textrm{r}=\left(N-N_0\right){/}\left(N+N_0\right)$. Then, the correction for the experimental coincidence signal is given by
	\begin{equation}
		c_\textrm{e}=\frac{\left(N-N_0\right)}{2}\left[ 1+v_\textrm{t}\cos(2\phi-\Phi_\textrm{h}) \right]+N_0,
		\label{eqce}
	\end{equation}
from where we can calculate the experimental visibility
	\begin{equation}
		v_\textrm{e}=v_\textrm{t}v_\textrm{e}^\textrm{r},\label{eqvis}
	\end{equation}
and, alternatively, write the experimental signal as
	\begin{equation}
		c_\textrm{e}=\frac{\left(N+N_0\right)}{2}\left[ 1+v_\textrm{e}\cos(2\phi-\Phi_\textrm{h}) \right],
	\end{equation}
with $0\leq v_\textrm{e}\leq\left(N-N_0\right){/}\left(N+N_0\right)$. 

The path-difference phase $\phi$ is introduced by rotating a mirror mounted on a motorised stage capable of rotations in steps of~$\sim10^{-5}$ degrees, see Fig. 2b in main text. Complete cycles ($\phi\to2\pi$) are obtained by physical rotations of the mirror of the order~$\sim 10^{-4}$ degrees.

The reference coincidence signal, i.e. for $\Phi_\textrm{h}{=}0$, was obtained by imposing $s{=}0$ and averaging all phase offsets by varying $\alpha$ between $\alpha{=}0$ and $\alpha{=}\pi$ in steps of $\Delta\alpha{=}\pi{/}20$. A maximum in the reference signal is equivalent to $\sim12,000$ coincidence events collected in $5$ seconds. We measure $\sim10$ kHz coincidences from the source, accordingly our set-up has  $\sim24\%$ overall transmission. The experimental errors arise from instabilities, e.g. small mechanical fluctuations in the mirror position ($<10^{-5}$ degrees), to which the Sagnac interferometer is sensitive to, in addition to inevitable random errors from Poissonian photon-counting statistics. Figure~{\ref{fig:fig1SI}} shows our recorded coincidence signals for four cases.

\subsection{IV. Measurement of $\Phi_{\textrm{h}}$ without Hong-Ou-Mandel interference}

	\begin{figure*}[htp!]
	\centerline{\includegraphics[width=0.85\textwidth]{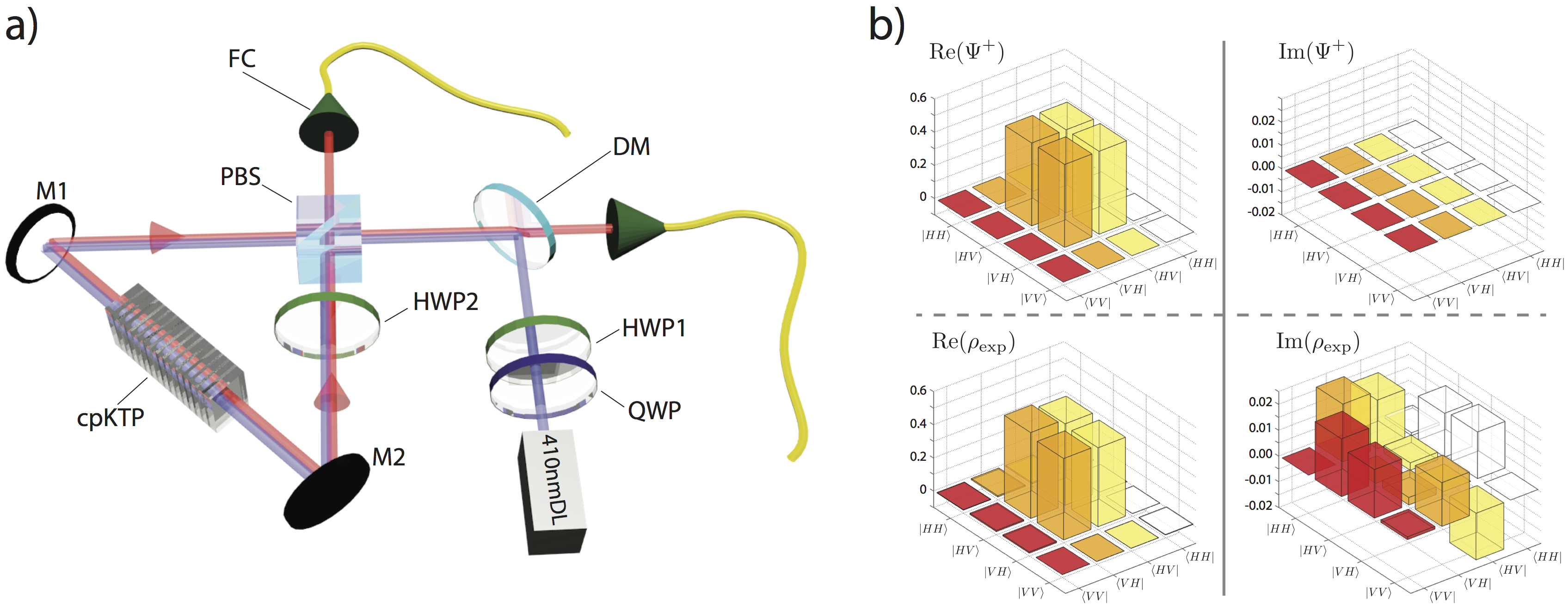}}
	\caption{{\bf Experimental depiction of two photon polarization entangled source.} a) A pair of polarization entangled single-photons are produced via spontaneous parametric downconversion in a $10$~mm long custom periodically-poled KTiOPO$_{4}$ (cpKTP) nonlinear crystal~\cite{Aggie2011}. A continuous wave pump centered at $410$~nm is output from a laser diode (410nmLD) with its polarization controlled by a half-wave (HWP1) and quarter-wave plate (QWP). The pump is incident on a dual-wavelength polarizing beam-splitter (PBS)~\cite{Smith:CQS} which together with mirrors M1 and M2 make up the Sagnac loop. The superposition of counter-propagating pump modes in the Sagnac loop destroys the `which path' information of the downconverted fields resulting in an entangled two-photon state that can be tuned from $\ket{HV}+e^{i\phi}\ket{VH}\rightarrow\ket{HV}$ by means of the HWP1 on the pump field. HWP2 is custom made to suit both the pump at $410$~nm and the downconverted light at $820$~nm, it works to flip the reflected vertically polarized pump mode from $\ket{V}\rightarrow\ket{H}$ for phase matching in the cpKTP crystal. A dichroric mirror (DM) and long pass filters (not shown) before each fibre coupler (FC) separate and block the pump light from being coupled to the output FCs. b) Experimental results from a quantum state tomography of the polarization entangled two-photon state $\rho_{\textrm{exp}}$. 
	}
	\label{fig:fig3SI}
	\end{figure*}
Although the presented data in the main text was taken with indistinguishable photons (inside the Hong-Ou-Mandel dip), qualitatively similar measurements were obtained with distinguishable photons by moving the modes outside of the dip, see Fig.~\ref{fig:fig2SI}. Given our particular evolutions however, it can be shown that the anti-bunching terms observed when there is a degree of distinguishability between the photons (absent in the completely indistinguishable case) do not alter the measured holonomic phases, $\Phi_{\textrm{h}}$. Instead, they show up merely as a background of counts, reducing the overall signal-to-noise ratio of the interference traces shown in Fig.~\ref{fig:fig2SI}. It was therefore not imperative to obtain perfect indistinguishability in order to measure the holonomic phases. Yet, any amount of indistinguishability will enhance the signal-to-noise ratio above that of the completely distinguishable case. A comparison between our measurements outside the Hong-Ou-Mandel dip (fully distinguishable) and the corresponding data inside the Hong-Ou-Mandel dip (indistinguishable), are shown in Fig.~{\ref{fig:fig2SI}}a-c.

\subsection{V. Source of polarization entangled photon pairs}

The experimental setup for producing two-photon polarization entangled states is shown in Fig.~\ref{fig:fig3SI}a). A state tomography is shown in Fig.~\ref{fig:fig3SI}b) and it reveals a tangle~\cite{Wooters:Conc} of $\mathbb{T}_\text{tomo}{=}0.959\pm0.001$ and a state fidelity of $\mathfrak{F}{=}0.988\pm0.001$ when compared to the state $\ket{\Psi^+}{=}\left(\ket{HV}+\ket{VH}\right){/}\sqrt{2}$.

%%%%%%%%%%%%%%%%%%%%%%%%%%%%%%%%%%%%%%%%%%%%%%%

\end{document}